\documentclass[pra,twocolumn,superscriptaddress,showpacs,floatfix]{revtex4}
\usepackage{graphicx}
\usepackage{amsmath}

\begin{document}

\title{Soliton-based discriminator of non-coherent optical pulses}
\date{\today}
\author{S.K. Turitsyn}
\author{S.A. Derevyanko}

\affiliation{Photonics Research Group, Aston University, \\ Aston Triangle,
Birmingham, B4 7ET, UK}

\begin{abstract}
We introduce a concept of noncoherent optical pulse discrimination from a
coherent (or partially coherent) signal of the same energy using a
phenomenon of soliton generation. The impact of randomisation of the optical
signal content on the observable characteristics of solitons generation is
examined and quantified for a particular example of rectangular pulse.
\end{abstract}


\maketitle

\section{Introduction}
The comparative study of the properties of coherent vs noncoherent
waves (fields, signals); and ordered vs disordered systems plays an
important role in both modern physics and technological applications
(see e.g. \cite{LGP,Mecozzi} and references therein). The examples
range from Anderson localization \cite{a58} to the interaction of
signals with noise the latter being one of the major factors
limiting performance of modern optical communication systems (see
e.g. \cite{Mecozzi}). Optically amplified transmission system
typically comprise a chain of optical amplifiers (e.g. erbium doped
fiber amplifiers) which add noise into the system. Amplifier noise
accumulated along the fiber links degrades the quality of
transmission. For instance, statistically rare occurrences of
parasitic noisy spikes having the same energy as the signal pulses
might lead to errors in the transmission of information.
The origin of such errors is in applying
sub-optimal receivers that
make decision based on information about the
 energy (or fraction of energy) of incoming pulse only and discard any information about the phase (as well as
 the internal structure of the pulse itself). Therefore there is no possible
 way for a standard detection device to distinguish between a true logical ``one''
 (characterised by high level of coherence) and the disordered spurious pulse of the same energy.
However, the latter ``ghost'' pulse generally differs from the true signal due to
a higher level of non-coherence and in this Letter we propose a soliton-based
device intended to separate the signal from random fluctuations of the same energy.
We also discuss here an interesting link between the problem of separation
of coherent signal component from non-coherent noise in fiber optics and a
theoretical problem of the emergence of localized states in a random
potential. In particular, we examine the generation of coherent localized
modes (solitons) from random initial field distribution (serving as a
potential in the corresponding non-Hermitian spectral problem).

Separation of a signal from noise is one of the fundamental research
problems that occurs in a broad range of scientific and technological
applications. Nonlinear techniques might be of a special interest in this
field as they can offer new opportunities that cannot be realized in linear
systems. For instance, nonlinear optical processing might enable one to
distinguish (in the optical domain) between the signal and noise accumulated
within the signal bandwidth - something which is not possible using linear
techniques \cite{hako}. Optical fiber nonlinearity is of particular interest
because the resulting nonlinear systems for some range of parameters, can be
described by well studied soliton models and powerful mathematical apparatus
can be applied to a variety of physical and technical problems (see e.g.\cite
{MG,hako,Mecozzi}). Note that optical soliton techniques are widely known in
the context of signal transmission. However their high potential in the
field of signal processing is yet to be explored.

Recent progress in advanced modulation formats \cite{Gna} and the growing
interest in coherent optical communications have emphasized the importance
of techniques that exploit the nature of the signal, in particular, the
optical phase. Signal processing utilising optical phase might offer many
advantages over traditional electrical processing that relies only on signal
intensity. Stable optical structures such as solitons might play a crucial
role in the design and development of a new generation of optical processing
devices that use signal phase as well as intensity. Recent demonstration of
the quasi-lossless fiber span \cite{PRL2006} that can be used as one of the
building blocks of optical soliton devices has rekindled interest in
soliton-based signal processing.

In this Letter we propose to use soliton generation as a technique to
discriminate between coherent (or almost coherent) signals and parasitic
non-coherent pulses of the same energy. We focus
here on the fiber-optic applications employing the on-off
keying data format - when logical ones are presented by pulses and logical zeros - by empty time slots.
We would like to point out, however, that the considered approach
can be generalised to a variety of systems and modulation formats.
Note that the propagation of partially coherent pulses in the
Kerr-type media has long been a subject of investigation in
different physical contexts, including random phase modulation of
temporal solitons \cite{bw86}, nonlinear Fraunhofer diffraction of
random fields \cite{AVC86,bkk} and other interesting examples
\cite{MS1}. We would also like to stress that here we
are not using solitons for transmission of information but
rather as building blocks of a nonlinear filter element which differs the proposed approach
from the one employed in conventional soliton-based transmission links.
 The reader interested in noise effects in conventional soliton transmission links is
 referred to the monograph \cite{Mecozzi} as well as the review \cite{HW96}.

\section{Problem statement}
An accumulation of the amplified spontaneous
emission noise along a fiber-optic link leads to the generation of randomly
modulated non-coherent background. In the on-off keying data format
transmission there exist time intervals when no deterministic signal is
present  (time slots corresponding to logical zeros). Radiative background
noise manifests itself as the noise generated ghost pulse. It fluctuates
and hence can achieve an energy high enough to be wrongly detected by a
receiver as logical one. Note that in the standard approach to signal
detection, the decision is made taking into account only the received pulse
energy, and any additional information related to the optical phase is not
used at all and is lost. Any noise field accumulated within the time
interval $T_{b}$ and having the optical bandwidth (optical filter bandwidth)
$B$ can be presented (following e.g. \cite{Mecozzi}) as a sum of
statistically independent modes:
\begin{equation}
q(t)=\sum_{k=-M/2}^{M/2}\,\eta _{k}\,\exp [i\,2\,\pi \,k\,t/T_{b}],\quad M=%
\mathrm{int}[B\,T_{b}]  \label{noise}
\end{equation}
here $\mathrm{int}[\ldots ]$ means the integer part of an
expression. The complex coefficients $\eta _{k}$ are independent
random variables with zero mean and variance $\langle \eta _{k}\eta
_{k^{\prime }}^{\ast }\rangle =(2\sigma ^{2}/T_{b})\,\delta
_{kk^{\prime }}$. For a chain of $N$ optical amplifiers with the
same gain coefficient $G$; the amplifier inversion factor $n_{sp}$;
and carrier frequency of the envelope $\omega _{0}$; the noise power
spectral density per polarization is $2\sigma ^{2}=n_{sp}\hbar
\,\omega _{0}\,(G-1)\,\,N\,$. \ The average energy of the noise is
$\langle E_{ASE}\rangle =2\sigma ^{2}\,(M+1)$, which is typically
small. However, some rare fluctuations can produce random spikes
with the energy close to that of a logical ``one'' degraded by the same
noise. It is highly desirable therefore to distinguish between noisy
(non-coherent) fluctuations and a deterministic localized pulse of
the same total energy $E$. Now by normalisation of
time using a bit length $T_b$ (see below) we can express the
normalised bandwidth through the value of $M$, therefore, in what
follows we will simply quantify the bandwidth by the number of
modes, $M$.

The key idea of the proposed approach is to use soliton generation effect
for the discrimination between coherent and noncoherent pulses.  Indeed, the
energy of a complex Gaussian field (\ref{noise}) is characterised by a $\chi ^{2}$ distribution with
$2(M+1)$ degrees of freedom, while the phase of each complex harmonic $\eta_k$
is uniformly distributed between 0 and $2\pi $. Such random phase and
amplitude modulation typical of parasitic pulses affects soliton generation
and, thus, can be detected through these changes in a passive fiber based
device. Without loss of generality, we consider here a true soliton system
that can be implemented in lossless fiber spans \cite{PRL2006} or in short
pieces of highly nonlinear fiber with negligible loss effect. Evolution in $%
z $ of the optical field $Q(t,z)$ is then governed by the well-known
Nonlinear Schr\"{o}dinger equation (NLSE) (\cite{hako,Mecozzi,MG}).

\begin{equation}
i\;\frac{\partial Q}{\partial z}+\frac{1}{2}\frac{\partial ^{2}Q}{\partial
t^{2}}+|Q|^{2}Q=0  \label{NLSE}
\end{equation}

Here Eq. (\ref{NLSE}) is written in dimensionless soliton units, with time
normalized by the pulse width, $t\rightarrow t/T$, and power normalized by $%
P_{s}=(\gamma L_{D})^{-1}\equiv |\beta _{2}|/(\gamma T^{2})$, $\gamma $ is
the nonlinear coefficient in W$^{-1}$/km, $L_{D}=T^{2}/|\beta _{2}|$ is the
dispersion length measured in km and $\beta _{2}$ is the (anomalous) group
velocity dispersion parameter measured in ps$^{2}$/km (see e.g. \cite{MG,Mecozzi}). Energy is measured in units $E_{s}=P_{s}T.$ The particular
shape of the input pulse in the proposed approach is not critically
important, but without loss of generality, we assume in what follows a real
rectangular pulse shape with width $T\leq T_{b}$, amplitude $Q_{0}$ and
energy $E=Q_{0}^{2}T$.

\section{Soliton generation.}
In the case of the real, unmodulated,
rectangular, initial condition; the energy threshold of the generation of $N$%
-th soliton can be readily derived \cite{m73}: $E_{N}=\pi ^{2}\,(2N-1)^{2}/4.
$ In the presence of noise, the deterministic formula for a number of
generated solitons $N=\mathrm{int}\left[ 0.5+\sqrt{E}/\pi \right] $ is
replaced by the set of probabilities $P_{N}$ of generating $N$ soliton states.
Those probabilities $P_{N}(E,\sigma )$ will depend on the pulse energy, $E$,
as well as the strength of disorder $\sigma $ (and the bandwidth parameter $M
$). Our main goal here is to determine how the dependence of an averaged
number of the generated solitons, $\langle N\rangle = \sum_{N=0}^\infty N\,
P_N$ on input pulse energy evolves when switching from a deterministic
localized input to a noisy non-coherent signal of the same energy.

Here it is pertinent to mention a closely related
problem of the influence of chirp (both deterministic and random) on
soliton generation \cite{bw86,dhal02,pdt07}. In particular, one may be interested in finding analytical criteria for the number of created solitons (and corresponding energy thresholds) for different initial pulse shapes. Indeed, when the  input pulse is a real function waveform a simple criterion by Kivshar and Burzlaff (KB) \cite{kb88} applies that relates the number of created solitons with the area of the pulse. In a particular case of the rectangular pulse one recovers the analytical result above. However, it is not straightforward to apply this criterion beyond the formal scope of its validity. Indeed, because this criterion relies only on the area of the pulse it is insensitive to any non-trivial phase modulation of the pulse. But any such phase modulation (deterministic or random) critically affects the soliton content of the pulse - the effect that is missed completely when applying KB criterion. For a simple case of a constant linear chirp the problem of soliton pulse generation from a Gaussian wave form was examined in  \cite{pdt07} where it was shown that linear chirp severely decreases the number of soliton states in the pulse. It is no surprise then that the KB criterion is not applicable also to random complex input (\ref{noise}). As an alternative, a standard WKB  approximation can be used to estimate the number of created solitons \cite{MNPZ} but its applicability
is restricted to the regime where the number of solitons is large. So unfortunately, there are no reliable analytical criteria for establishing the number of emerging solitons for the output in the form (\ref{noise}).

\subsection{Zakharov-Shabat spectral problem for random potentials.} According
to the inverse scattering transform theory any initial condition of the NLSE
(\ref{NLSE}) evolves into a combination of solitons plus quasi-linear
oscillating wave packets. The parameters of formed solitons, as well as
their number, can be derived via the non self-adjoint Zakharov-Shabat
spectral problem (ZSSP) \cite{MNPZ} which is a special case of more general
Ablowitz-Newell-Kaup-Segur scheme (see  \cite{akns73}):
\begin{equation}
\left\{
\begin{array}{cc}
i\,\partial \psi _{1}/\partial t+Q\,\psi _{2} & =\zeta \psi _{1} \\
-i\,\partial \psi _{2}/\partial t+R\,\psi _{1} & =\zeta \psi _{2}
\end{array}
\right.  \label{akns}
\end{equation}
with complex potentials $R$ and $Q$. The ZSSP spectral problem for Eq.(\ref
{NLSE}) corresponds to the non self-adjoint reduction of (\ref{akns}): $%
R\equiv -Q^{\ast }$. Other important reductions include $R\equiv Q^{\ast }$
which corresponds to NLSE with positive (normal) group velocity dispersion
(or defocusing nonlinearity) and the case $R\equiv 1$ that leads directly to
the Hermitian Schr\"{o}dinger equation with the ``energy'' given by $\zeta
^{2}$ connected with the Korteveg-de-Vries equation. In any reduction, the
random nature of the potential leads to interesting properties of the
spectra. In particular, the reduction which yields the linear
Schr\"{o}dinger equation with random potential is known to exhibit \textit{%
Anderson localization} \cite{a58,LGP}, which manifests itself as an
exponential decay of the transmission coefficient through a disordered
potential with the growth of the width of the potential, $T$. The
corresponding decrement $\tau (\zeta )$ is known as \textit{localization
length} (or rather, localization time in our notations). For the
self-adjoint reduction $R=Q^{\ast }$, a similar phenomenon exists \cite
{gky90} where it was proven that the localization length does not depend on
the spectral parameter $\zeta $. In what follows we will focus on the
focusing NLSE (\ref{NLSE}) with non-selfadjoint reduction where each
(random) discrete eigenavlue located in the upper complex half-plane of $%
\zeta$ corresponds to a bright optical soliton emerging from a noisy input
pulse. As mentioned in the previous section the existing deterministic criteria for soliton creation are not applicable to the noncoherent complex pulses of the form (\ref{noise}).  Therefore we have to resort to the numerical Monte Carlo (MC) simulations of the stochastic  ZSSP eigenproblem (\ref{akns}) and the problem of determination of the average number of solitons,   $\langle N\rangle $ reduces thus to the calculation of the distribution and the average number of discrete eigenvalues of ZSSP.

\section{Concept of soliton discriminator and results}
Now we present the results of the MC simulations for different system parameters. Those are to be compared with the analytical result (as discussed above) for a
deterministic rectangular pulse of the same energy. Again we would like to
convey the idea that we are specially interested in rare fluctuations when
the energy of the ``ghost pulse'' coincides with that of a real pulse. In
the MC simulation we had to insure that for each realisation of the input
field the total energy was fixed at the same level, $E$. This was achieved
by picking $(M+1)$ complex Fourier harmonics with equal amplitudes (fixed by
the total energy $E$) and independent random phases uniformly distributed
between $0$ and $2\pi $. Figure 1 shows the average number of emerging
solitons for totally non-coherent input (\ref{noise}) versus pulse energy
(normalised by the soliton energy), for different number of random
harmonics in log-log scale. It is seen from Fig. 1 that the introduction of  randomness dramatically reduces the probability of soliton generation; and the average numbers of solitons can be less then unity, even for relatively high values of energy. This demonstrates a feasibility of the proposed concept of soliton discriminator since it shows that a totally non-coherent input pulse in unlikely to produce a stable soliton content. An interesting question remains open of whether the different curves in Fig.1 have a finite size scaling (FSS) with $M$, i.e. whether there exists an effective ``phase
transition'' in $M$. In other words, whether there exist a universal function $f$ and parameters $\alpha$, $\beta$ and $E_{th}$ so that the average number of solitons scales with energy as $M^{\alpha} \,f((E-E_{th})/M^{\beta})$. Standard algorithms \cite{bs01} have failed to establish such a scaling, but the results are still inconclusive and more advanced analysis is needed.

\begin{figure}[h]
\label{fig:1} \includegraphics[width=8.6cm]{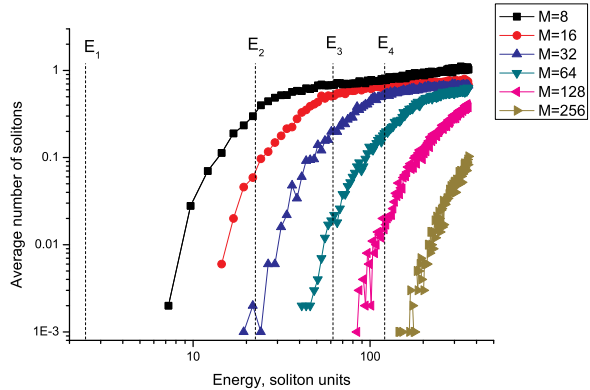}
\caption{(Color online) The average number of emerging solitons versus pulse
energy for non-coherent input pulse for different number of modes M.
Vertical lines show deterministic thresholds ($E_{1}=2.46,\,E_{2}=22.46,%
\,E_{3}=61.8,\,$and so on) of N-soliton generation in soliton units.}
\end{figure}
\begin{figure}[h]
\includegraphics[width=8.6cm]{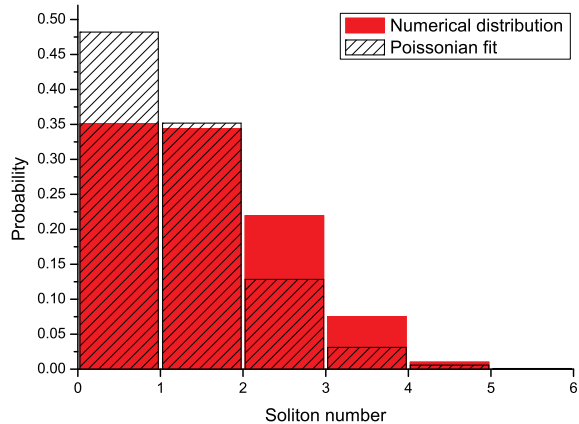}
\caption{(Color online) The probability distribution of the number of created solitons (solid red bars) and the Poissonian fit (dashed bars).}
\end{figure}
\begin{figure*}[h!]
\label{fig:2} \includegraphics[width=17.8cm]{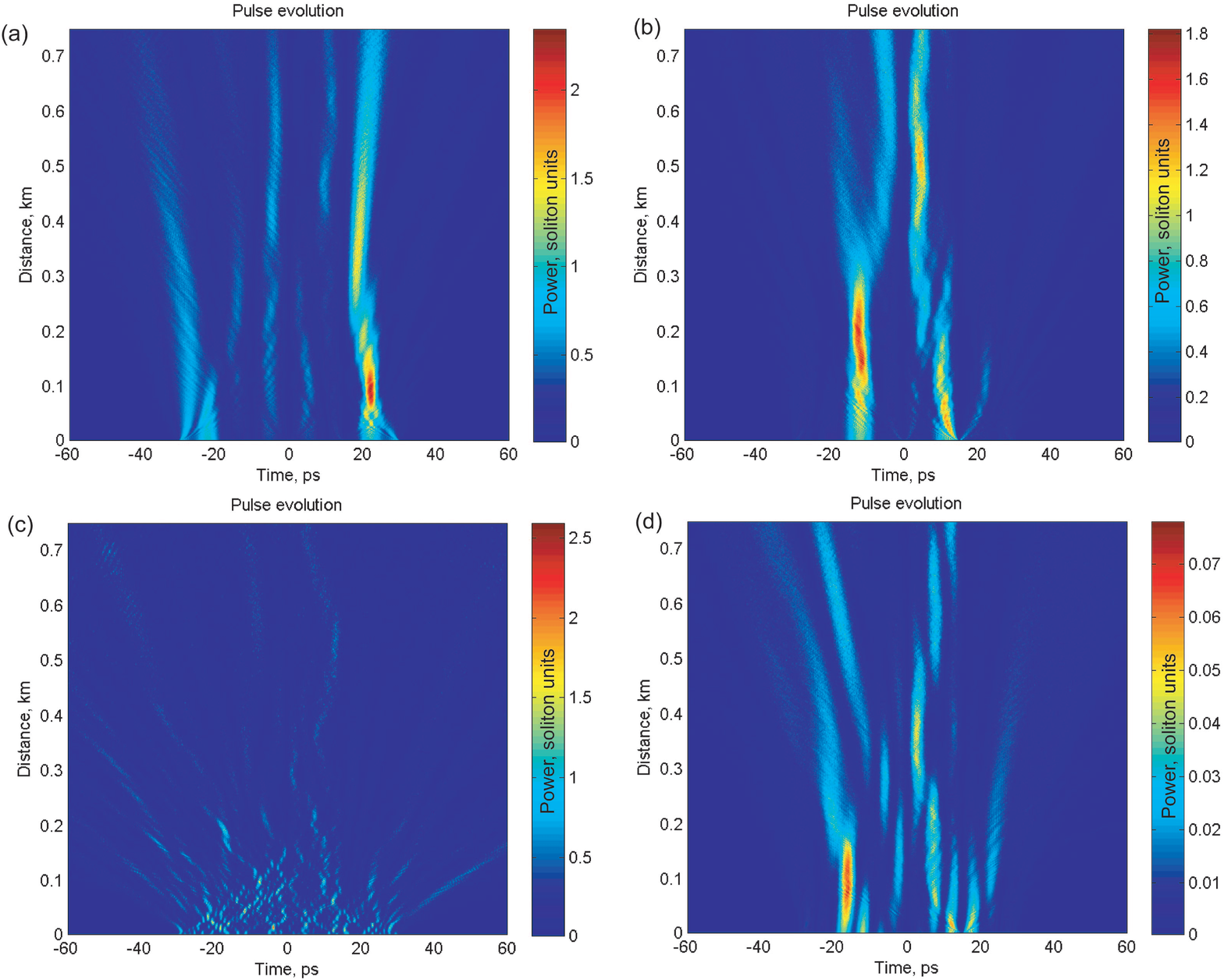}
\caption{(Color online) Possible scenarios of stochastic soliton creation. The energy (in soliton units) and the number of modes are: $E=360$, $M=8$ for (a) and (b), $E=360$, $M=64$ for (c) and $E=13.5$, $M=16$ for (d).}
\end{figure*}

Another observation is that the deterministic thresholds of soliton creation (shown in Fig.1 as vertical dashed lines) also do not apply for the disordered input. We define a soliton creation threshold for a disordered pulse as the first value of $E$ where the average value of solitons is non-zero. The computationally resolved thresholds for different numbers of harmonics are given by the left end points at the graphs in Fig.1. As seen from Fig.1 the data points in the vicinity of the thresholds tend to fluctuate which is the natural consequence  of the Monte-Carlo averaging procedure: the first non-zero contribution to the average number of solitons  come from the extremely rare fluctuations that are notoriously difficult to sample. In our simulations  we saw no solitons created below the first deterministic threshold $E_1=2.46$ after 1000 Monte Carlo runs. Because FSS algorithm failed to establish the unique threshold energy $E_{th}$ we cannot rigorously prove  that such processes are forbidden. This issue requires a separate study with higher resolution and advanced   Monte-Carlo techniques (like e.g. multicanonical sampling). This however lies beyond the scope of the current paper.

Fig.1 shows only the average number of solitons and
not the corresponding probability distributions of soliton formation
$P_N$. It is instructive to study the individual probabilities
$P_N$, and in particular verify whether they follow a Poissonian
distribution. Therefore, in Fig. 2 we also present separately the
probability distribution obtained via Monte Carlo simulation of a system with $M=8$ noisy harmonics and energy $E=360$ (in soliton units).  The results are compared with the Poissonian fit with the same average $\langle N \rangle = 0.73$. One can see that the numerical distribution is somewhat wider and less localized than the Poissonian fit.

Let us now illustrate different possible scenarios of a ``ghost pulse'' propagation by providing space-time traces of pulse evolution for different values of parameters $E$ and $M$. The goal of this detour is to help the reader to gain a qualitative understanding of ``dangerous" noise fluctuations that are able to trigger the creation of spurious solitons during pulse evolution.  In Fig. 3 we provide space-time traces of pulse evolution from the stochastic input (\ref{noise}). We opted here for the following system parameters: $T_b=60$ ps, $\beta_2=-20$ ps$^2$/km, $\gamma=3$ W$^{-1}$ km$^{-1}$.  The four scenarios shown in Fig. 3 are: one soliton generation at high energy and low number of noisy modes (Fig. 3(a)), two soliton generation for the same parameters (Fig. 3(b)), no soliton created for high energy and high number of  modes (Fig. 3(c)) and, finally, no soliton generated at sub-threshold low energy (Fig. 3(d)).

So far we have only considered the evolution and soliton content of a purely disordered pulse of the form (\ref{noise}).
However typically in optical communications a signal arriving at the
decision device is only partially non-coherent and generally can be
presented as a deterministic component $Q_{0}(t)$ mixed with an accumulated
noise $q(t) $, $Q(t)=Q_{0}(t)+q(t)$ where $q(t)$ is given by (\ref{noise}).
For this case we compute the number of emerging soliton states for the ZSSP (%
\ref{akns}) for the input pulse $Q(t)$ rather then for $q(t)$. Figure 4
shows the impact of noise on the generation of solitons in this case. The
contribution of noise to the initial pulse (i.e. the measure of
non-coherence) is characterised by an optical signal-to-noise ratio, $%
OSNR=E/(2\sigma ^{2}\,(M+1))$ a quantity often used in optical
communications. Figure 4 demonstrates two important limits: large OSNR corresponds
to the deterministic limit described analytically above, and the limit of
small OSNR leads to the situation of fully non-coherent pulse similar to
Fig. 1. It also gives a qualitative estimate of the level of non-coherent
noisy content in the pulse that can be distinguished by such a nonlinear
soliton-based filter. It is seen from Fig. 4 that pulses with OSNR less than
roughly 10 dB have quite different characteristics in terms of the number of
generated solitons compared to their deterministic counterparts of the same
energy (top of the picture). This means that if the non-coherent share of
the total pulse energy content is more than approximately one tenth, the
proposed device should be able to discriminate between such partially
coherent signals and non-coherent noisy ghost pulses.
\begin{figure}[h]
\label{fig:3} \includegraphics[width=8.6cm]{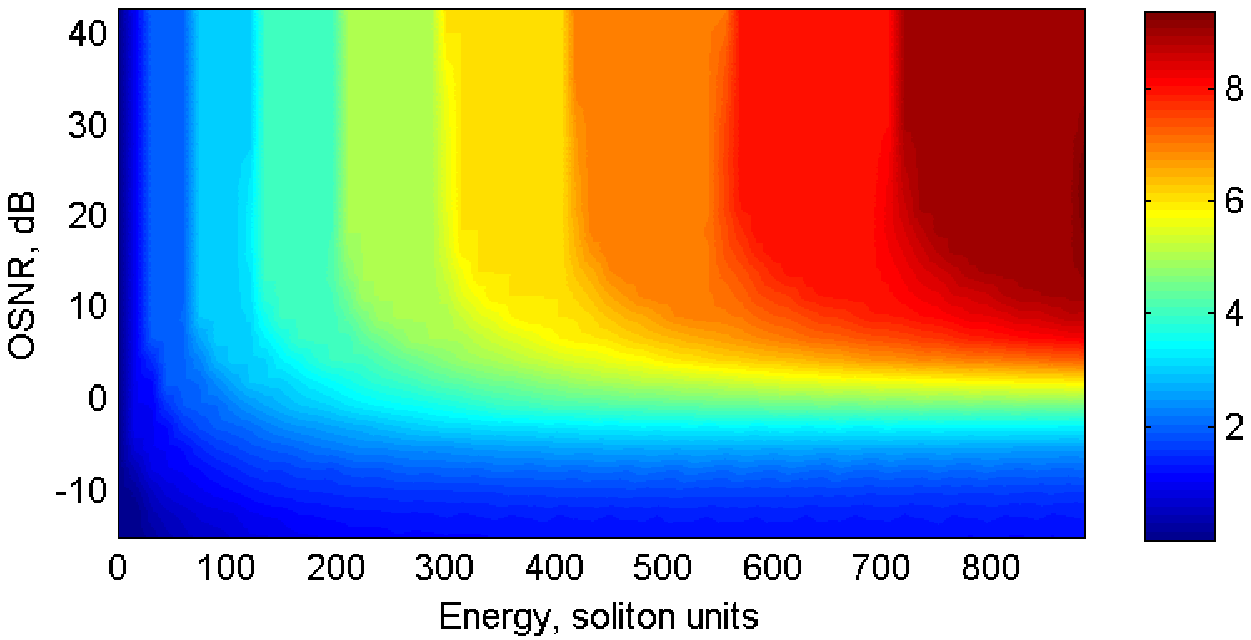}
\caption{(Color online) The averaged number of solitons versus the pulse
energy $E$ and OSNR$.$ Here M = 64.}
\end{figure}

\begin{figure}[h]
\label{fig:4} \includegraphics[width=8.6cm]{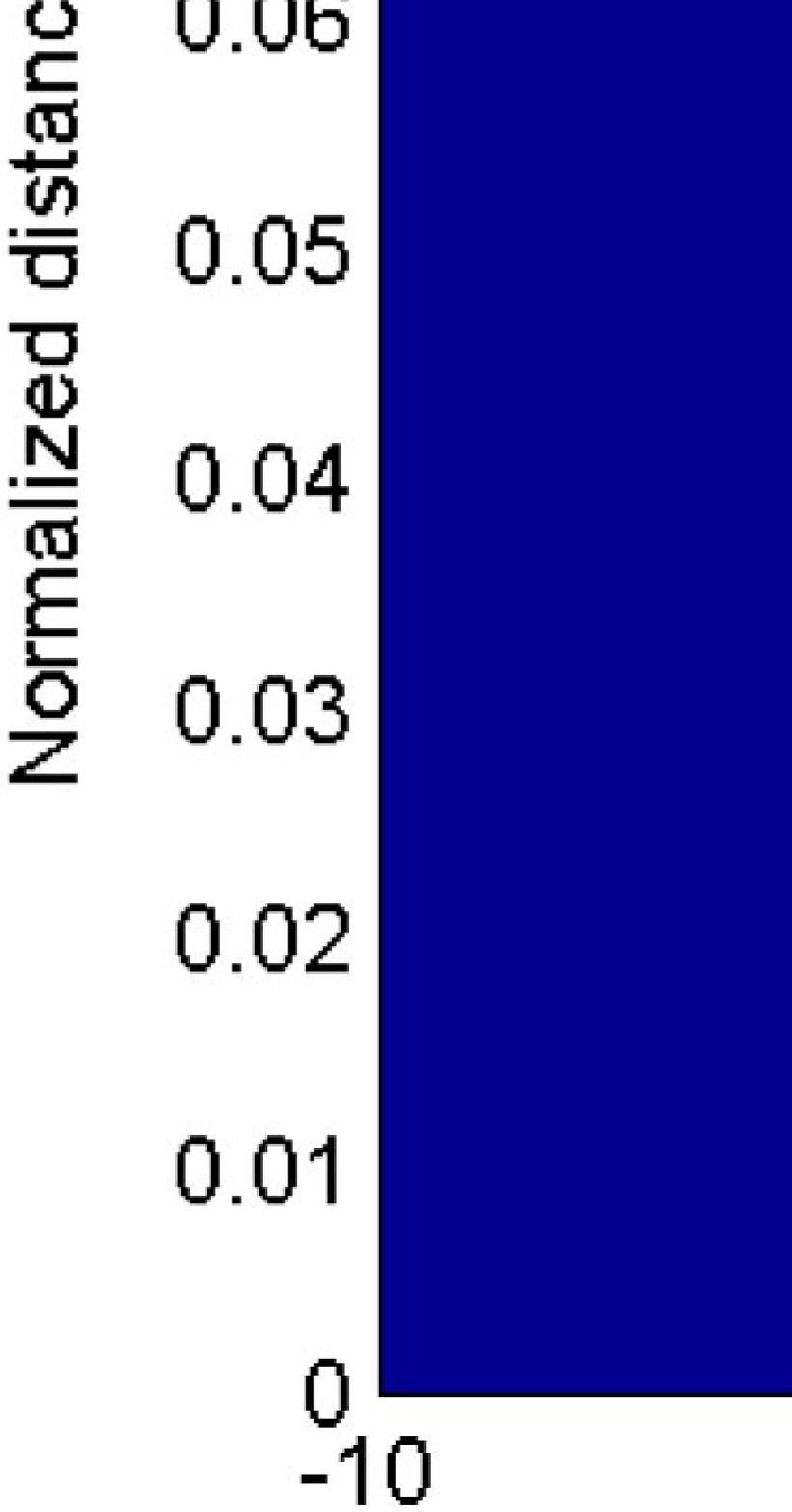}
\caption{(Color online) The evolution of a given pattern with the real
central ``one'' (a) and a spurious central ``one'' (b) one of the same
energy }
\end{figure}

The action of a soliton discriminator is illustrated in Fig. 5 where a
typical evolution of a given pattern 10100 is shown in the two cases: (a)
when all bits are deterministic and (b) when the central bit is a noisy,
non-coherent ghost pulse of the same energy as a deterministic one in (a).
For the particular  realization of the noisy input (bottom panel) we picked $%
M=64$ harmonics with the same total energy as the coherent pulse in (a).
Figure 5 demonstrates how non-coherence prevents the creation of soliton
structures, and one readily observes at the top panel (a) of the figure that
most of the pulse energy quickly settles in an emerging single soliton which
is reverse to the case (b) at the bottom panel. In the real world units this
discrimination can be observed at the propagation scales of several
kilometers for a standard fiber or just several meters for highly nonlinear
fibers. This implies that a potential soliton discriminating device will not
not require large fiber spans and could thus be quite portable. We would
like to emphasize that Fig. 5(b) presents just a single realization of the
event when a noisy background accumulates energy comparable to that of the
logical one. In other words the proposed soliton discriminator might be
particulary efficient when the system performance is limited by such
fluctuations. Figure 5 illustrates that the proposed device is feasible and
is capable of discriminating between coherent and non-coherent inputs even
if the energies are the same and no difference would have been detected by a
standard integrate-and-dump receiver.

Let us discuss now an issue of the speed of
convergence of a disordered input to a final (multi)soliton state.
Indeed, when analysing a general pulse-to-soliton conversion problem
there are quite a few examples when such a convergence to a final
sech-shape  solution is very slow (see e.g. \cite{kns95,ch1,ch2}).
This, however, does not have a serious impact on the proposed concept of soliton discriminator. Indeed, in order for the soliton discriminator to operate one does not have to wait until the amplitude of the emerged soliton stabilises at a certain fixed level. It is not even required at all for the solution to stabilise at a prescribed ``clean'' sech-shape. The important point is to have a solution with a finite energy after the radiation has been shed - this will ensure that true logical ``ones'' survive the transmission. The final shape of the emerging solution is not of importance for this application, it just has to be localized in time domain. In the  meantime it is desirable that a pulse of the same energy, but generated from random Fourier harmonics  -- Eq.(\ref{noise}) -- will simply collapse and will not give rise to any soliton  state at all. And Fig. 5 illustrates just that: a ``ghost one'' collapses into linear radiation very quickly (and therefore will rightly be detected as ``zero'') while the true logical ``one'' evolves into an asymptotic soliton state. True, the convergence in the latter case is slow, but the most important fact is that it survives and still carries enough energy to be rightly detected as ``one''.

Evidently, there is a number of technical issues to be resolved before any
practical implementation of the proposed soliton discriminating device can
be achieved. In particular, timing jitter of partially non-coherent signals
should be accounted for in the analysis of errors. Also pattern effects and
soliton interaction must be considered in any practical implementation.
However there is an important factor that makes the concept of a soliton
discriminator quite feasible. All the adverse effects listed are more
pronounced when the propagation distance increases (e.g. timing jitter) and
cannot be neglected when propagating through distances in the order of a
couple of dispersion lengths $L_{D}$. However, as stressed earlier the
soliton discriminator can operate at much smaller distances. A distance of
several kilometers or even meters usually suffices thus diminishing the
majority of mentioned detrimental effects. In addition, a number of standard
well developed control techniques (modulation, optical filtering) can be
applied to reduce these effects even further.

\section{Conclusions}
We have proposed soliton generation as a candidate
technique to discriminate between coherent signals and parasitic
non-coherent (or partially coherent) pulses of the same energy. The
principal idea was to use lossless transmission spans (modelled by
integrable NLSE) to filter noisy dispersive part of the solution from the
emerging soliton part. Using numerical modelling we have demonstrated the
feasibility of the proposed approach. We have also determined how the
average number of the generated solitons is affected by the presence of
noise in the partially non-coherent input signal. This work was supported by
the Royal Society and the EPSRC.

\end{document}